\documentclass[apj]{emulateapj}
\usepackage{graphicx}
\usepackage{graphics}
\usepackage{subfigure}
\usepackage{natbib}

\begin{document}

\title{An Empirical Measure of the Rate of White Dwarf Cooling in 47 Tucanae}

\author{ R.~Goldsbury\altaffilmark{1}, J.~Heyl\altaffilmark{1}, H.~B.~Richer\altaffilmark{1}, P.~Bergeron\altaffilmark{2}, A.~Dotter\altaffilmark{3}, J.~S.~Kalirai\altaffilmark{3,4}, J.~MacDonald\altaffilmark{5}, R.~M.~Rich\altaffilmark{6}, P.~B.~Stetson\altaffilmark{7}, P.-E.~Tremblay\altaffilmark{8}, K.~A.~Woodley\altaffilmark{1}}

\altaffiltext{1}{Department of Physics \& Astronomy, University of British Columbia, Vancouver, BC, Canada V6T 1Z1; rgoldsb@phas.ubc.ca, heyl@phas.ubc.ca, richer@astro.ubc.ca, kwoodley@phas.ubc.ca}
\altaffiltext{2}{D\'epartement de Physique, Universit\'e de Montr\'eal, C.P.~6128,
Succ.~Centre-Ville, Montr\'eal, Qu\'ebec H3C 3J7, Canada; bergeron@astro.umontreal.ca}
\altaffiltext{3}{Space Telescope Science Institute, 3700 San Martin Drive, Baltimore, MD, 21218; dotter@stsci.edu, jkalirai@stsci.edu}
\altaffiltext{4}{Center for Astrophysical Sciences, Johns Hopkins University, Baltimore, MD, 21218}
\altaffiltext{5}{Department of Physics \& Astronomy, University of Delaware, Newark, DE, 19716; jimmacd@udel.edu}
\altaffiltext{6}{Division of Astronomy and Astrophysics, University of California at Los Angeles, Los Angeles, CA, 90095;  rmr@astro.ucla.edu}
\altaffiltext{7}{National Research Council, Herzberg Institute of Astrophysics, Victoria, BC, Canada V9E 2E7; peter.stetson@nrc-cnrc.gc.ca}
\altaffiltext{8}{Zentrum f{\"u}r Astronomie der Universit{\"a}t Heidelberg, Landessternwarte, K{\"o}nigstuhl 12, 69117 Heidelberg, Germany; ptremblay@lsw.uni-heidelberg.de}

\begin{abstract}

We present an empirical determination of the white dwarf cooling sequence in the globular cluster 47 Tucanae.  Using spectral models, we determine temperatures for 887 objects from Wide Field Camera 3 data, as well as 292 objects from data taken with the Advanced Camera for Surveys.  We make the assumption that the rate of white dwarf formation in the cluster is constant.  Stellar evolution models are then used to determine the rate at which objects are leaving the main sequence, which must be the same as the rate at which objects are arriving on the white dwarf sequence in our field.  The result is an empirically derived relation between temperature ($T_{eff}$) and time ($t$) on the white dwarf cooling sequence.  Comparing this result to theoretical cooling models, we find general agreement with the expected slopes between 20,000K and 30,000K and between 6,000K and 20,000K, but the transition to the Mestel cooling rate of $T_{eff} \propto t^{-0.4}$ is found to occur at hotter temperatures, and more abruptly than is predicted by any of these models.

\end{abstract} 

%\maketitle

\section{Introduction}

The majority of stars that form in our Galaxy will eventually end up as white dwarfs.  Given enough mass to fuse hydrogen but not enough mass to eventually form a neutron star or a black hole, the range of initial masses for white dwarf stars encompasses much of the lower end of the initial mass function.  The study of such objects can therefore reveal a wealth of information about the population that formed them.  A predictive model of white dwarf cooling can also be used to derive precise ages for stellar populations \citep{oswalt-1996-nat,hansen-2007-apj}.  Additionally, white dwarfs themselves can serve as laboratories for constraining the physics that determine their cooling rates.

The aim of this work is to obtain an empirical measurement of the rate at which white dwarfs cool.  Theoretical work in this field began with Mestel's 1952 paper in which he layed out a simple model for the rate at which white dwarfs should lose energy.  This model assumes a completely degenerate and isothermal core surrounded by a thin non-degenerate envelope.  Thermal motion of the non-degenerate ions as well as the slow collapse of the star itself releases gravothermal energy from the core of the white dwarf, while the rate of energy loss from the star is determined by the opacity in the envelope \citep{mestel-1952-mnras}.  The end result is a prediction that the temperature of a white dwarf should decrease with time since formation, $t$, as $t^{-0.4}$.  This basic model has been gradually improved upon since then, with the introduction of more and more physical considerations.  Summaries of progress in this field can be found in \citet{dantona-1990-araa}, \citet{hansen-2004-pr}, and most recently \citet{althaus-2010-aa}.  Hansen also presents a comparison of some modern cooling models.  However, comparisons of white dwarf cooling models to observable populations tend to focus on the cool end of the sequence, which is the most relevant for dating.  We present here a detailed comparison of a number of current models to measurable cooling rates up to temperatures of 30,000K.  

In this paper, we will analyze a group of white dwarfs that have all evolved from the same initial population.  These objects are part of the Galactic globular cluster 47 Tucanae (NGC 104).  All of the objects in this cluster formed at the same time, or at least formed over a period of time that is very small when compared with the current age of the cluster.  The fact that our entire sample of white dwarfs comes from a single population allows us to draw a number of simplifying assumptions.  

The first is that all of the white dwarfs are of a similar mass.  Because all of the stars should have formed around the same time, all of the objects on the main sequence have had roughly the same amount of time to evolve.  This implies that the variance in the masses of objects currently leaving the main sequence should be small.  Initial final mass relations \citep{kalirai-2009-apj} then suggest that variance in the mass of the white dwarfs, which these turn-off stars will eventually become, is also small.  This agrees with the analysis of white dwarf spectra in globular clusters, which suggest that all of the objects are of similar mass \citep{kalirai-2009-apj,moehler-2004-aa}.  It is important to note that this assumption will not be true for very old white dwarfs, but since our entire sample should be younger than a few Gyrs, this assumption is valid.  

The second assumption is that objects arrive on the white dwarf sequence at a constant rate.  This follows from the fact that the rate at which objects are leaving the main sequence is constant.  This is supported by a comparison of the luminosity function of the lower giant branch with a model luminosity function with a constant evolution rate (see Section \ref{timescale}).  Given that this rate is constant, and all objects leaving the main sequence will eventually become white dwarfs, the rate at which objects are arriving onto the white dwarf sequence must also be constant.  

These two points allow us to put all of our white dwarfs on the same cooling sequence by fitting models to each object's spectral energy distribution (SED) to determine a temperature, and then requiring that the white dwarfs are uniformly distributed in time along the sequence.  This approach allows us to measure the rate at which the temperature of the white dwarfs is changing in a way that is independent of any white dwarf cooling models.

\section{Data}

Our data set consists of 121 orbits of Hubble Space Telescope (HST) observations.  During these orbits, exposures were taken simultaneously using the Advanced Camera for Surveys (ACS) and Wide Field Camera 3 (WFC3).  The current paper focuses on the WFC3 UVIS data in $F390W$ and $F606W$.  These data are comprised of 26 orbits covering 13 adjacent fields (1 orbit in each filter).  The description of the methodology in this paper will refer to that data set.  Similar methods were used to produce a cooling curve from the ACS data, which is taken in $F606W$ and $F814W$.  The WFC3 data cover over 60 arcmin$^2$ over a range of 6.5-17.9 pc from the core of 47 Tuc.  The observations and data reduction are described in detail in \citet{kalirai-2012-aj}.

\section{Selecting our White Dwarf Sample}

\begin{figure}[h!tbp]
\centering
\includegraphics[scale=0.85]{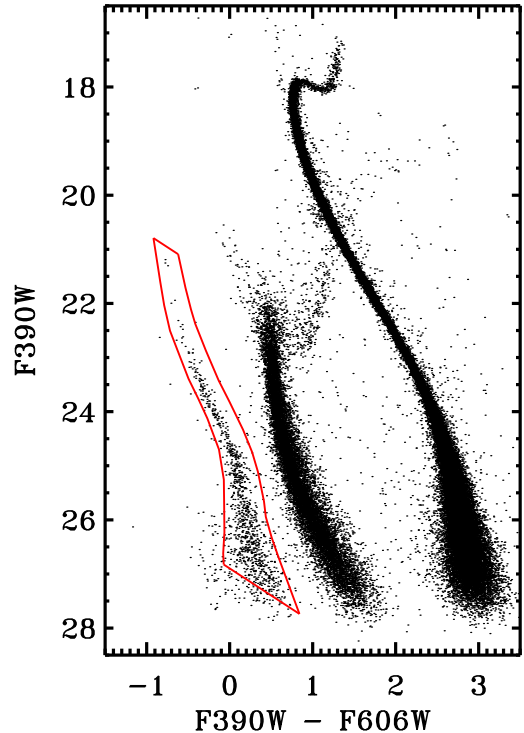}
\caption{The CMD after making selection cuts in error, sharpness, and $\chi^2$ of the PSF fit.  The final WD selection region is shown in the enclosed area (red).  887 objects are found within this enclosed region.}
\label{cutcomp}
\end{figure}

Selection began with a catalog generated from the final drizzled images as described in \citet{kalirai-2012-aj}. RMS residuals of the PSF fit from DAOPHOT II \citep{stetson-1987-pasp}, SHARP from ALLSTAR \citep{stetson-1994-pasp}, and photometric errors from DAOPHOT were used to cut down the full catalog to a higher quality sample.  Stars that made the cut were required to have errors of less than 0.1 magnitudes in $F390W$ and $F606W$, a SHARP value within $2\sigma$ of the mean SHARP value, and an RMS of the PSF fit less than 2.5\%.  The CMD after selection cuts are made is shown in Figure \ref{cutcomp}.

The region around the white dwarf sequence between roughly 22nd and 27th magnitudes in F606W was bounded manually.  The objects inside this region were used as our final sample.  Although some real objects are almost certainly lost due to these cuts, this can be properly accounted for during the analysis of artificial star tests.  The same cuts are used to determine which objects are ``found" from our input artificial stars.

Uncertainty also arises from manually bounding our sequence.  We analyze the distribution in color along the white dwarf sequence and model the spread in color as Gaussian. If we integrate this Gaussian over our enclosed area and compare this to the total integral,  we find that we expect to lose $14$ objects due to this bounding region.  This means that of our final sample of 887, each object should count as (14+887)/887 objects (prior to incompleteness corrections).  The uncertainty introduced by this is of order 1.5\% in the time scaling of our object list.  This scaling is discussed in more detail in Section \ref{timescale}.

\section{Fitting White Dwarf Temperatures}
\label{fit_temp}

To determine an effective surface temperature ($T_{eff}$) for each object in our sample, we utilize spectral models from \citet{Tremblay-2011-apj}.  The models are for thick atmosphere (the H envelope mass fraction is approximately $q_H=10^{-4}$) DA white dwarfs, and are parametrized over a grid of surface gravity and $T_{eff}$ values ranging from 6.0 to 10.0 and 6,000 K to 120,000 K, respectively.  The DB fraction is assumed to be negligible, which is supported by analysis from \citet{woodley-2012-aj}. These models ($F_{mod}$) are then combined with the filter throughput curves ($S_{\lambda}$), which are available online from \citet{stsci-2009-web}, to generate a magnitude in a particular filter for a given set of model parameters.  This is shown in equation \ref{eqn:specmag}.

\small

\begin{equation}
\label{eqn:specmag}
M_{\mathrm{mod}}=-2.5\log_{10}\!\left(\frac{\int_0^\infty \! \lambda F_{\mathrm{mod}} E_{\lambda} S_{\lambda} \, \mathrm{d}\lambda}{\int_0^\infty \! \lambda F_{0} S_{\lambda} \, \mathrm{d}\lambda}\right)
+5\log_{10}\!\left(\frac{d}{R}\right)
\end{equation}

\normalsize

\begin{figure}[bp]
\centering
\includegraphics[scale=0.5]{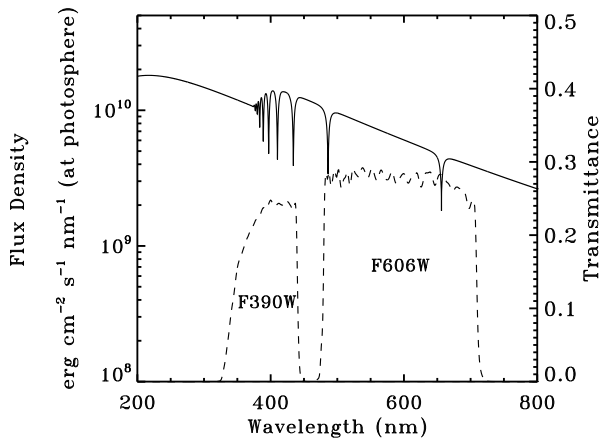}
\caption{A model WD spectrum with an effective surface temperature of 10,000K and a surface gravity of log(g)=7.5.  Throughput curves for our two filters are shown overplotted on a separate axis scale.}
\label{filtplot}
\end{figure}

The model magnitudes are fit to our sample of 887 white dwarfs using a maximum likelihood fitting method.  The likelihood of the data given the model is:

\small

\begin{equation}
\label{eqn:likelihood1}
L(data\mid parameters)=\prod_{i=1}^2\mathrm{exp}\!\left[-\frac{(data_{i}-mod_{i})^{2}}{2\sigma_{i}^{2}}\right]
\end{equation}

\normalsize

\begin{figure}[h!tbp]
\centering
\includegraphics[scale=0.5]{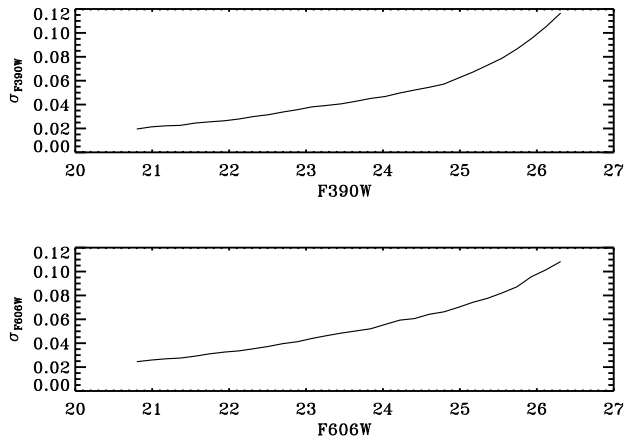}
\caption{The standard deviation of input - output magnitude derived from artificial star tests described in \citet{kalirai-2012-aj} (for all fields combined).}
\label{fig:errdist}
\end{figure}

Here the model parameters are $(m-M)_0$, $E(B-V)$, $M_{WD}$, and $T_{eff}$.  To condense this to the likelihood of $T_{eff}$ given our data, we integrate $L(data\mid parameters)$ over $(m-M)_0$, $E(B-V)$, and $M_{WD}$ with Gaussian priors of $13.30 \pm 0.15$, $0.04 \pm 0.02$, and $0.53 \pm 0.02$ M$_{\odot}$ respectively \citep{woodley-2012-aj}.  The distance prior used here is the mean of all values quoted in \citet{woodley-2012-aj} that do not depend on the white dwarf sequence.  The maximum likelihood in $T_{eff}$ is found for each object.  The uncertainty in the distance is twice as large as the value quoted by \citet{woodley-2012-aj}.  This more accurately reflects the distribution of distance measurements made by various groups, as presented in that paper.  The $\sigma_{i}$ in Equation \ref{eqn:likelihood1} is the estimated photometric error of each data point.  These values are derived from artificial star tests presented in \citet{kalirai-2012-aj}.  The error distributions of the two WFC3 filters are shown in Figure \ref{fig:errdist}.

\section{Determining the Timescale}
\label{timescale}

\subsection{A Sorted List of Temperatures}

After an effective surface temperature is determined from fitting spectral models to the photometry of each object, this list must be sorted from hottest to coolest.  This sorted list is the first step toward determining the rate at which the white dwarfs are cooling.  Because the objects are arriving on the white dwarf sequence at a constant rate, the position in the sorted list should be proportional to the time spent on the sequence.  That is to say, the time between when the 100th and 200th objects arrive on the sequence should be the same as the time between the 200th and 300th.

Completeness of the sample also needs to be taken into account.  The cooler objects are also fainter and thus less likely to be seen in a crowded field.  Using artificial star tests as described in \citet{kalirai-2012-aj}, we can assign a completeness correction to each object by first binning the artificial star data by magnitude and position.  The completeness correction or fraction in some bin is then:

\begin{equation}
C_{corr}=\frac{N_{input}}{N_{found}} \ \ \ \mathrm{  or  } \ \ \ C_{frac}=\frac{N_{found}}{N_{input}}
\end{equation}

This can be interpolated to the magnitude of each object in our sample and the distance from the cluster center.  To calculate the position of each object in a completeness corrected sorted list we use the following equation:

\begin{equation}
x(n)=\sum\limits_{i=0}^n C_{corr}(i).
\label{xn_eq}
\end{equation}

\begin{figure}[h!tbp]
\centering
\includegraphics[scale=0.5]{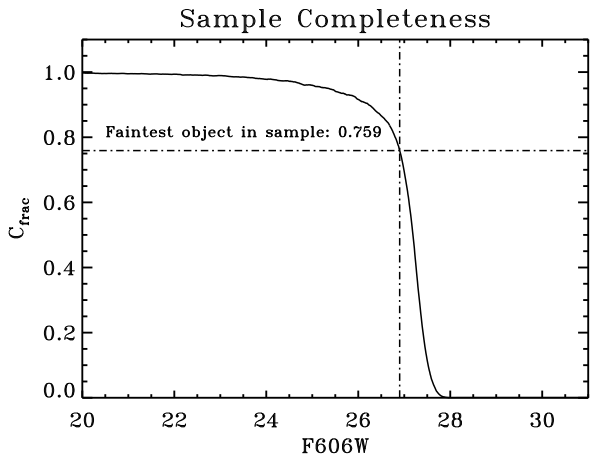}
\caption{Completeness fraction of our WD sample as a function of F606W magnitude (averaged over all fields).}
\label{compfig}
\end{figure}

It is easy to see that if the entire sample is 100\% complete, then this is just an array of positions in the sorted list (1, 2, 3, etc).  In our sample the completeness corrections are small.  Even the faintest objects are more than 75\% complete, and overall the sample is 89\% complete.  The completeness fraction as a function of magnitude is shown in Figure \ref{compfig}.  The artificial star tests were actually analyzed field by field.  Fields closer to the center of the cluster are obviously more dense, and have larger incompleteness.  The results in Figure \ref{compfig} are the total for all fields.  Incompleteness was also considered in both filters.  It was found to make no difference whether incompletenes values from a single filter, or a combination of both filters was used.  Since the corrections were so small, this had no effect on our final fitted parameters.  Completeness corrections for the ACS data were calculated in the same way.  The entire ACS sample is greater than 90\% complete.  

\begin{figure}[h!tbp]
\centering
\subfigure{
\includegraphics[scale=0.515]{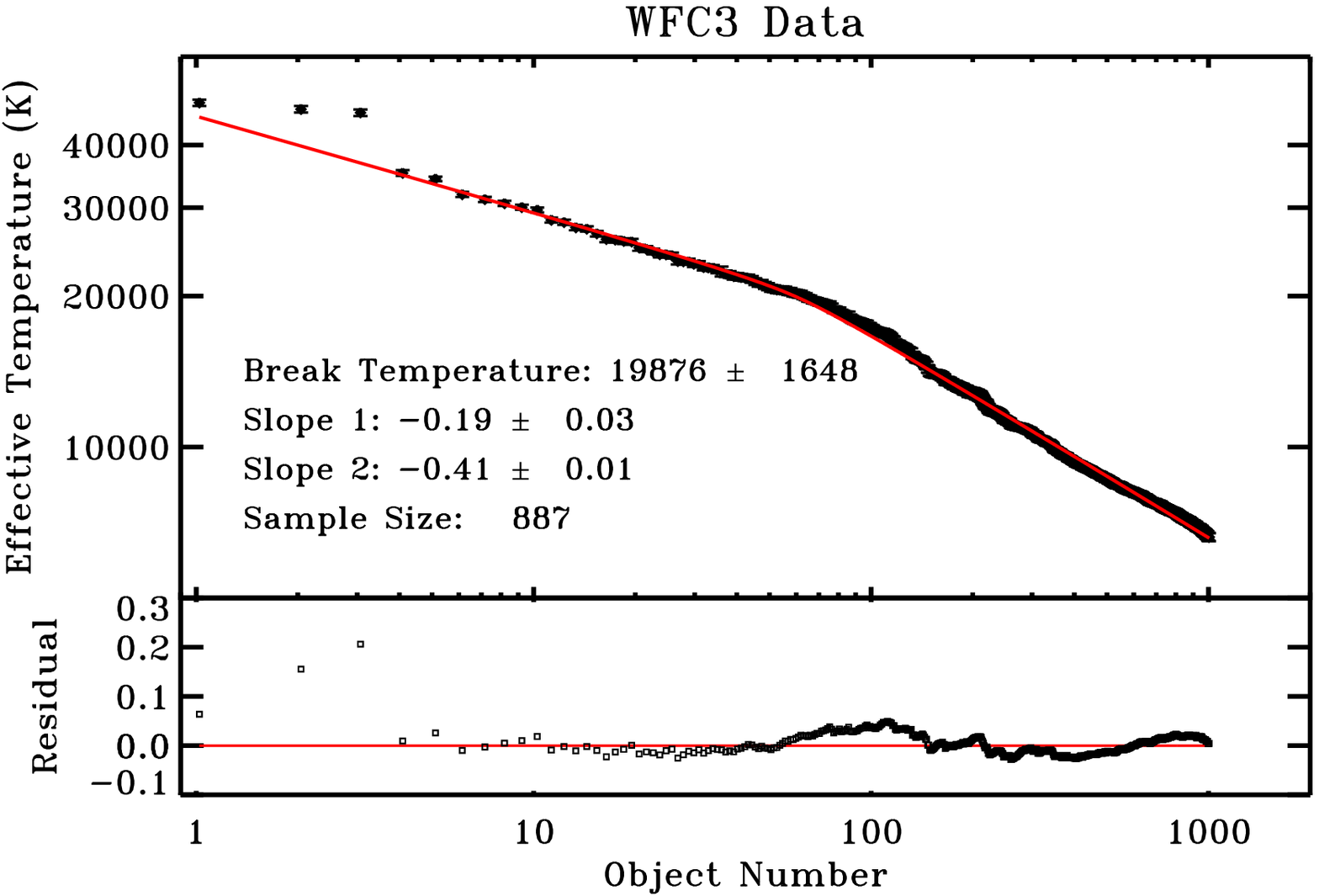}
}
\vspace{12pt}
\subfigure{
\includegraphics[scale=0.5]{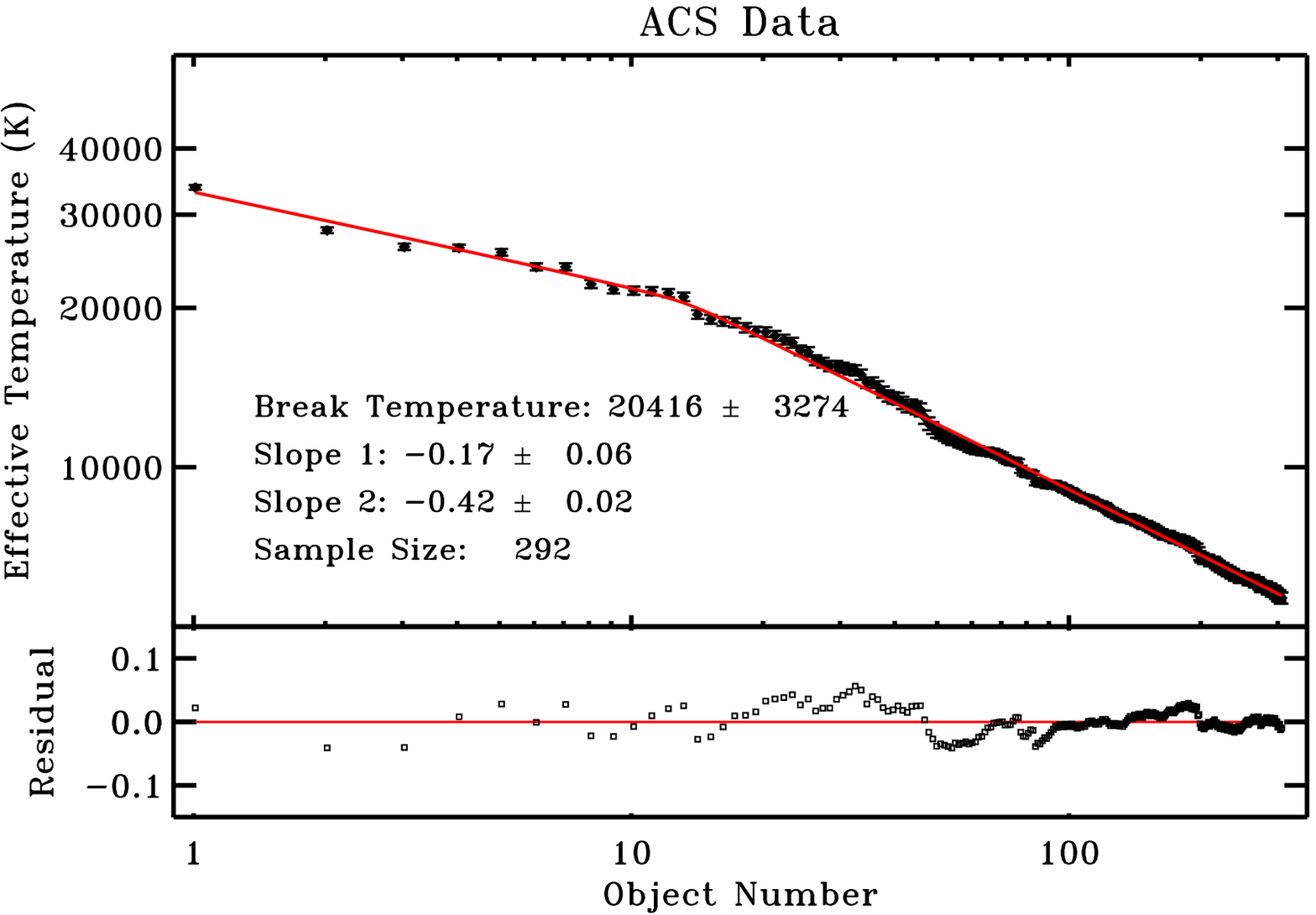}
}
\caption{The fitted $T_{eff}$ as a function of object number for our WDs.  The object number is proportional to the time on the white dwarf sequence, and is calculated as shown in equation \ref{xn_eq}.  The y-axis is just the temperature fit to each object's SED, as discussed in section \ref{fit_temp}.  The results from both data sets agree well within uncertainties.  Residuals are shown as $\frac{Data-Model}{Data}$ (the fractional residual).}
\label{listplot}
\end{figure}

A sorted, completeness corrected list is shown in log-log space in Figure \ref{listplot}.  This relation is well fit with a broken power law.  The slopes and ``break" of such a fit are also shown in the Figure.  These parameters do not depend at all on the absolute scaling of the x-axis, since this only acts as an additive constant in log-space (log$_{10}\left[\Delta t / object \right]$).  These power law parameters and their uncertainties were derived by iterating our temperature fitting analysis.  In each iteration we assume parameters of (m-M)$_0$, E(B-V), and M$_{WD}$ with values drawn from the prior distributions discussed above.  After each iteration, the temperatures are sorted and fit with this broken power law.  The distributions of these fit parameters define the best fitting value, and the uncertainties.  This process was also repeated for the ACS field with 292 WDs.  The results are shown in the lower panel of Figure \ref{listplot}.  All fitted parameters are consistent with the parameters from the WFC3 data.  The uncertainties in these fit parameters take into account the uncertainty in the photometry of our individual objects,  as well as the uncertainties in all of our marginalized parameters.

\subsection{The Rate at Which Stars are Leaving the Main Sequence}
\label{msrate}

\begin{figure}[h!tbp]
\centering
\includegraphics[scale=0.5]{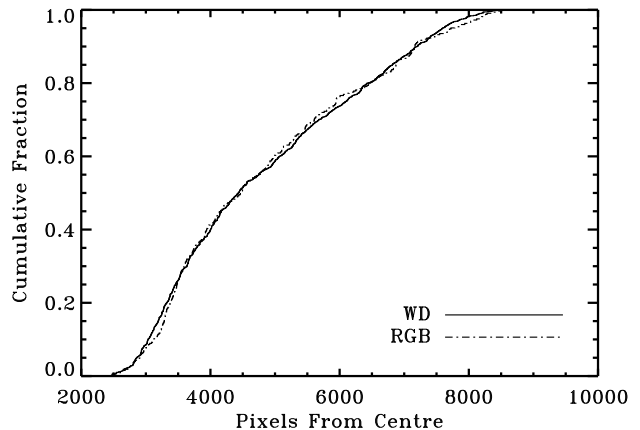}
\caption{A comparison of the cumulative radial distributions of the WD and lower RGB stars.  A KS test indicates that there is a 73\% chance that two samples of this size drawn from the same underlying distribution would differ by at least this much.  We therefore find no evidence of significantly different radial distributions for the WD and RGB stars.}
\label{radfig}
\end{figure}

In order to put the completeness corrected list into units of time, so that our cooling curve can be compared with theoretical models in both dimensions, we need to find the rate at which objects are arriving onto the white dwarf sequence in our field.  This rate should be the same as the rate at which objects are leaving the main sequence.  One could imagine that dynamical relaxation of white dwarfs could cause problems with this assumption.  The white dwarfs are considerably less massive than their giant branch progenitors and should be expected to relax on timescales similar to the relaxation time for the field.  However, this relaxation time is long.  At the half-light radius the relaxation time is $3.5$ Gyrs \citep{harris-1996-aj}.  Using the velocity dispersion profile of 47 Tuc calculated by \citet{lane-2010-mnras}, and the density profile from our own data, we calculate an expected mean relaxation time for our WFC3 field to be 10 Gyrs.  We can compare this to the expected time since the white dwarfs in our sample were the mass of current turn-off stars, which is $< 2.5$ Gyrs for stars brighter than 27.8 in $F390W$ (utilizing models from \citet{fontaine-2001-pasp}).  Given that these objects have only existed for a small fraction of a relaxation time at their current mass, there should be no observable mass segregation.  Additionally, comparing the radial distributions of the white dwarfs and the giant branch stars, no significant difference is apparent, as shown in Figure \ref{radfig}.  This indicates that the white dwarfs have not yet relaxed, and their distribution in the field should be the same as their progenitors.  To determine the age scaling for our cooling curve we need to analyze the giant branch of the CMD.  The upper giant branch is saturated in our exposures, so we use the catalog from \cite{stetson-2000-pasp} of an overlapping field in B and V, selecting only those objects that would be found within the region of our WFC3 observations.

\begin{figure}[h!tbp]
\centering
\includegraphics[scale=0.8]{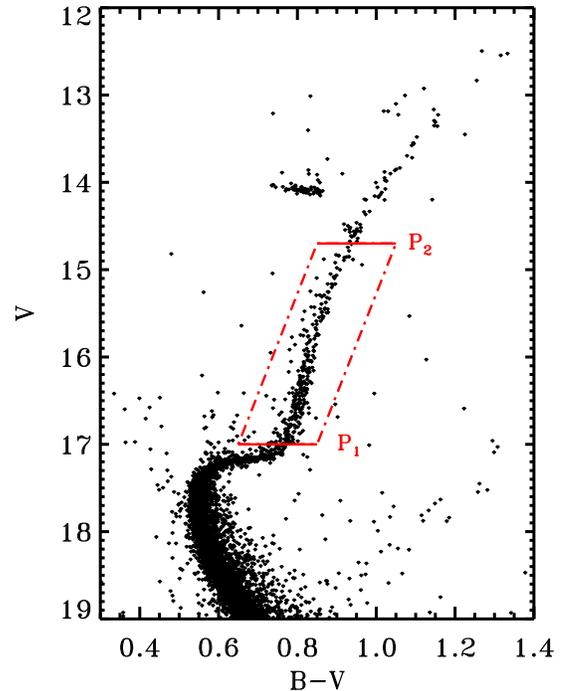}
\caption{The CMD from \cite{stetson-2000-pasp}, including only objects that fall within the region of our WFC3 data.  The region of the CMD between points $P_1$ and $P_2$ is fit with models from \cite{dotter-2008-apjs} to determine the time it takes an object to evolve between the points.}
\label{sgbplot}
\end{figure}

On this portion of the CMD, stars evolve almost vertically as their outer layers expand accompanied by a very moderate decrease in effective surface temperature.  To determine an evolution rate from these objects, we use theoretical models from \citet{dotter-2008-apjs}.  Our input parameters for the model are [Fe/H], (m-M)$_0$, current cluster age ($A$), and an age range ($\Delta A$).  This age range corresponds to the time it takes an object to evolve from one point on the lower giant branch to another point (point $P_1$ to point $P_2$ in Figure \ref{sgbplot}).  In our case, all of these parameters except the age range are well determined.  We calculate the likelihood of our magnitude range resulting from this model over this whole parameter space using Equation \ref{agelikeeq}.  We marginalize out all parameters except the age range, using priors that describe the measured values of [Fe/H], (m-M)$_0$, and $A$.

\begin{equation}
L(\Delta M_r\ |\ parameters)=\mathrm{exp}\!\left[\frac{-(\Delta M_m-\Delta M_r)^2}{2\sigma^2}\right]
\label{agelikeeq}
\end{equation}

\noindent $\Delta M_r$ is the selected magnitude range (shown in Figure \ref{sgbplot}), while $\Delta M_m$ is the range predicted by the model given a particular choice of input parameters.  \citet{chaboyer-2002-apj} show that there is an uncertainty of order 3\% in the age of such models.  To first order, object age and magnitude are linearly related in this region, and so this translates to an uncertainty of 3\% in the output magnitude range.  This uncertainty is used as $\sigma$ in Equation \ref{agelikeeq}.  The priors used in Equation \ref{agelikeeq} are all Gaussian with the following means and standard deviations: $[Fe/H]=-0.75\pm0.08$, $(m-M)_{F606W}=13.42\pm0.15$, and $A=11.5\pm0.75$ Gyr \citep{mcwilliam-2008-apj,woodley-2012-aj,dotter-2009-iau}.

\vspace{12pt}

\begin{figure}[h!tbp]
\centering
\includegraphics[scale=0.5]{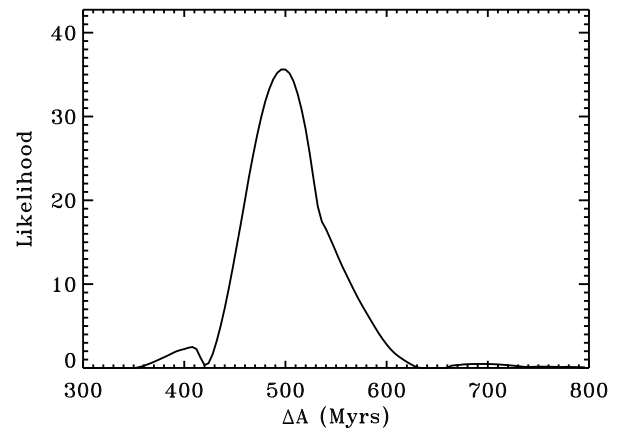}
\caption{Likelihood of $\Delta A$ fit to our selected magnitude range, after marginalizing out all other model parameters.  $\Delta A$ corresponds to the time it takes an object to evolve from point $P_1$ to point $P_2$ in Figure \ref{sgbplot}.}
\label{modlike}
\end{figure}

The result is a likelihood distribution of $\Delta A$ for our chosen magnitude range that takes into consideration uncertainties in the input model parameters, as well as uncertainty in the model itself.  This distribution is shown in Figure \ref{modlike}.

\begin{figure}[h!tbp]
\centering
\includegraphics[scale=0.5]{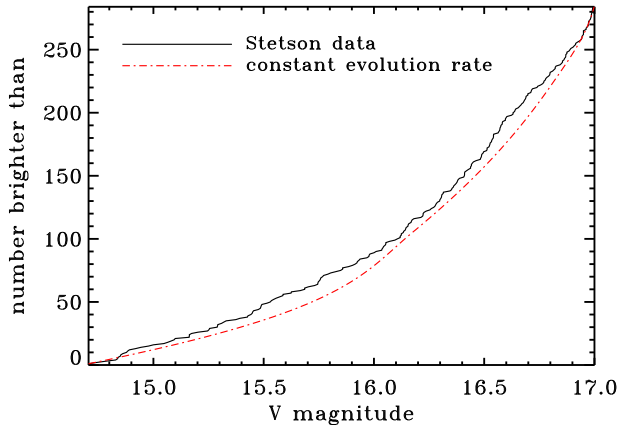}
\caption{A cumulative distribution of Stetson's RGB magnitudes between $P_1$ and $P_2$ in Figure \ref{sgbplot} compared to the expected distribution assuming a constant evolution rate.  A KS test suggests that there is a $40\%$ chance that a sample drawn from the model distribution would differ by at least as much as this data does.}
\label{stetline}
\end{figure}

Our model prediction for the timespan corresponding to this chosen magnitude range is $496_{-83}^{+95}$ Myrs (2$\sigma$ uncertainties).  The uncertainties on this value are entirely due to uncertainties in the model as well as the input parameters.  To determine the rate at which objects are evolving off the main sequence, the fit time interval must be divided by the number of objects ($N=287$) in this region of the CMD (shown in Figure \ref{sgbplot}) within our field.  The uncertainty in the number of objects must be $\sqrt{N}$, resulting in a final rate of 1.7 $\pm$ 0.2 Myrs per object, an age scaling uncertainty of ~12\%.  This value has no physical meaning for the cluster as a whole.  It is dependant entirely on our field.  However, as this is the same field as our WD sample, it is appropriate to assume that this rate is the same as the rate at which objects are arriving on the WD sequence.  This time/object is then multiplied by our ordered and completeness corrected list, resulting in an empirically measured relation between the temperature of the white dwarfs, and their age on the sequence.  This method also assumes that the rate of evolution off the main sequence is constant.  This assumption is well supported by analyzing the cumulative distribution of magnitudes between the two points in Figure \ref{sgbplot}.  This distribution is shown compared against a model distribution assuming a constant evolution rate in Figure \ref{stetline}.

\section{The Cooling Curve}

\begin{figure*}[!h!tbp]
\centering
\includegraphics[scale=0.95]{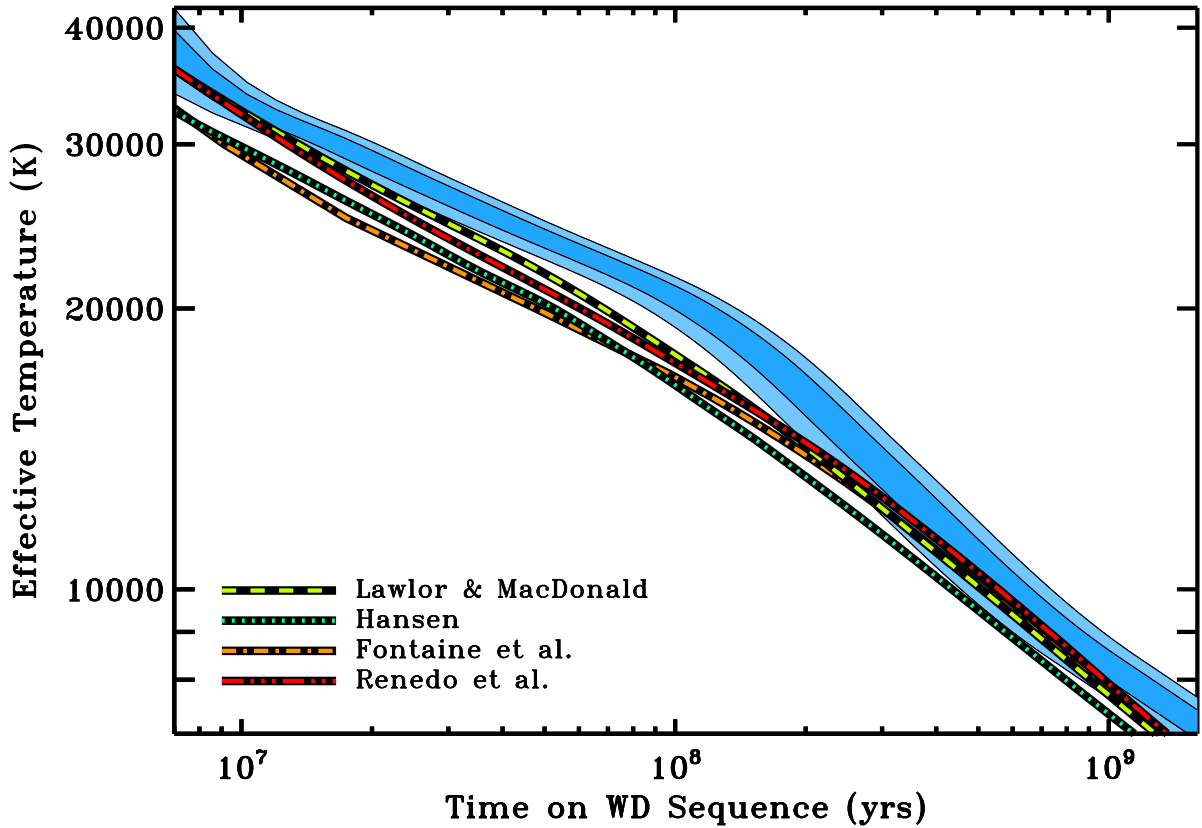}
\caption{The measured one and two sigma uncertainties of our cooling curve (the shaded blue regions).  The shape in this region is well constrained as shown in Figure \ref{listplot}, but the age scaling leads to a large systematic uncertainty in the x-direction.  This is the region represented by these contours.  Models of white dwarf cooling rates from \citet{fontaine-2001-pasp}, \citet{macdonald-2006-mnras}, \citet{hansen-2007-apj}, and \citet{renedo-2010-apj} are shown for comparison.}
\label{contcool}
\end{figure*}

The cooling curve determined from our data is shown in Figure \ref{contcool}.  The shape of the cooling curve is well constrained, as can be seen in the residuals of Figure \ref{listplot}.  The uncertainties quoted there represent the $1 \sigma$ uncertainties in the upper and lower slopes taking into consideration both the random uncertainties in the magnitudes of each object, as well as the uncertainty in the magnitude calibration \citep{kalirai-2012-aj}.  The contours shown in Figure \ref{contcool} display the systematic uncertainty in the x-direction that arises from our uncertain time-scaling factor.  This is the dominant source of uncertainty in our time scaled cooling curve.  

\subsection{The Models}

We compare our data to models from four groups.  The models of \citet{fontaine-2001-pasp} have pure C cores with H and He envelope mass fractions of $q_H=10^{-4}$ and $q_{He}=10^{-2}$.  The \citet{hansen-2007-apj} models also have envelope mass fractions of $q_H=10^{-4}$ and $q_{He}=10^{-2}$, and a mixed C/O core with profiles from \citet{salaris-1997-apj}.  Models from \citet{macdonald-2006-mnras} are evolved from the main sequence.  By the time they reach the white dwarf stage, the envelope mass fractions are $q_H=9.5\times10^{-5}$ and $q_{He}=3.1\times10^{-2}$.  The core is uniformly mixed with C and O mass fractions of 0.1 and 0.9 respectively.  Lastly, we use the models of \citet{renedo-2010-apj}, specifically those made publicly available with Z=0.001.  The envelope thickness of these models varies with time and is shown in Figure 4 of \citet{renedo-2010-apj}.  The core composition is discussed in section 3.6 of that same paper with the C/O mass fractions being roughly 0.3 and 0.7.

\subsection{Model Comparison}

As can be seen in Figure \ref{contcool}, the general shape of the empirical cooling curve agrees reasonably well with all four models.  The models all show a linear region with a slope of approximately $-0.20$ followed by a transition region, and then a section that agrees with Mestel's value of $-0.4$.  From our simple broken power law fit we find that the centre of our power law break occurs at an age of 104 $\pm$ 26 Myrs, and a temperature of 19900 $\pm$ 1600 K.  Fitting the same broken power law to the models yields transition ages of 190 Myrs (Fontaine et al.), 129 Myrs (Lawlor \& MacDonald), 107 Myrs (Hansen), and 250 Myrs (Renedo et al.).  This makes the time of our power law transition consistent with both Lawlor \& MacDonald and Hansen within 2$\sigma$, but inconsistent with Fontaine et al. and Renedo et al. at greater than 3$\sigma$.  However, the temperatures at which this transition occurs in the models seems to differ from our data.  These are 14,100K (Fontaine et al.), 16,600K (Lawlor \& MacDonald), 16,200K (Hansen), and 13,500K (Renedo et al.).  These differ at around 2.5$\sigma$ for Lawlor \& MacDonald and Hansen, and close to 4$\sigma$ for Fontaine et al. and Renedo et al.  It is difficult to draw conclusions about how well these models fit the data by condensing our entire cooling curve down into a few statistics.  A more robust method involves comparing our observed cumulative distribution of temperatures to those that would be predicted by the models.  In such an analysis, the scale of the x-axis is entirely irrelevant since only the relative time spent at a given temperature compared to any other temperature will determine the shape of the cumulative distribution.  This approach also compares the distribution over the entire temperature range, rather than condensing our curve down to a few statistics, such as the location of the power law break.

\begin{figure}[h!tbp]
\centering
\includegraphics[scale=0.5]{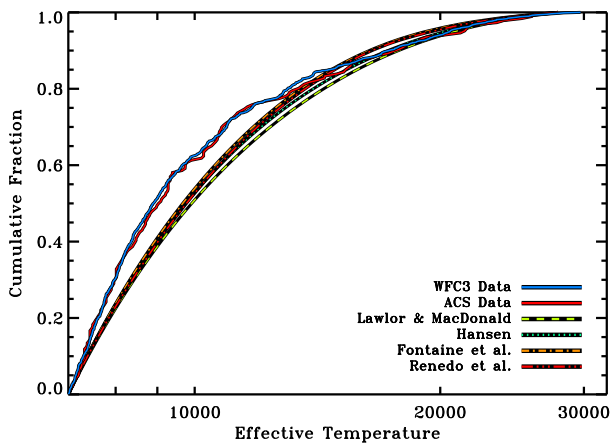}
\caption{This plot shows measured cumulative distribution of our objects found to have temperatures between 7,000 K and 30,000 K (813 objects WFC3, 188 objects ACS), as well as those predicted by the four models shown in Figure \ref{contcool}.}
\label{contks}
\end{figure}

\begin{figure}[h!tbp]
\centering
\includegraphics[scale=0.5]{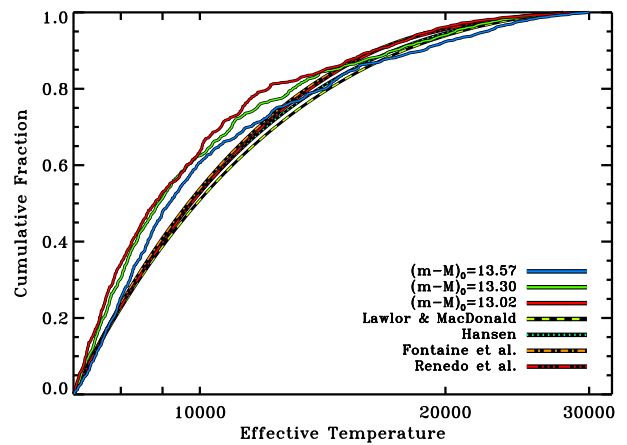}
\caption{Cumulative distribution of our fit temperatures compared to the models when assuming values at the extreme ends of literature distance measurements for 47 Tuc.}
\label{ks_dist}
\end{figure}

\begin{figure}[h!tbp]
\centering
\includegraphics[scale=0.95]{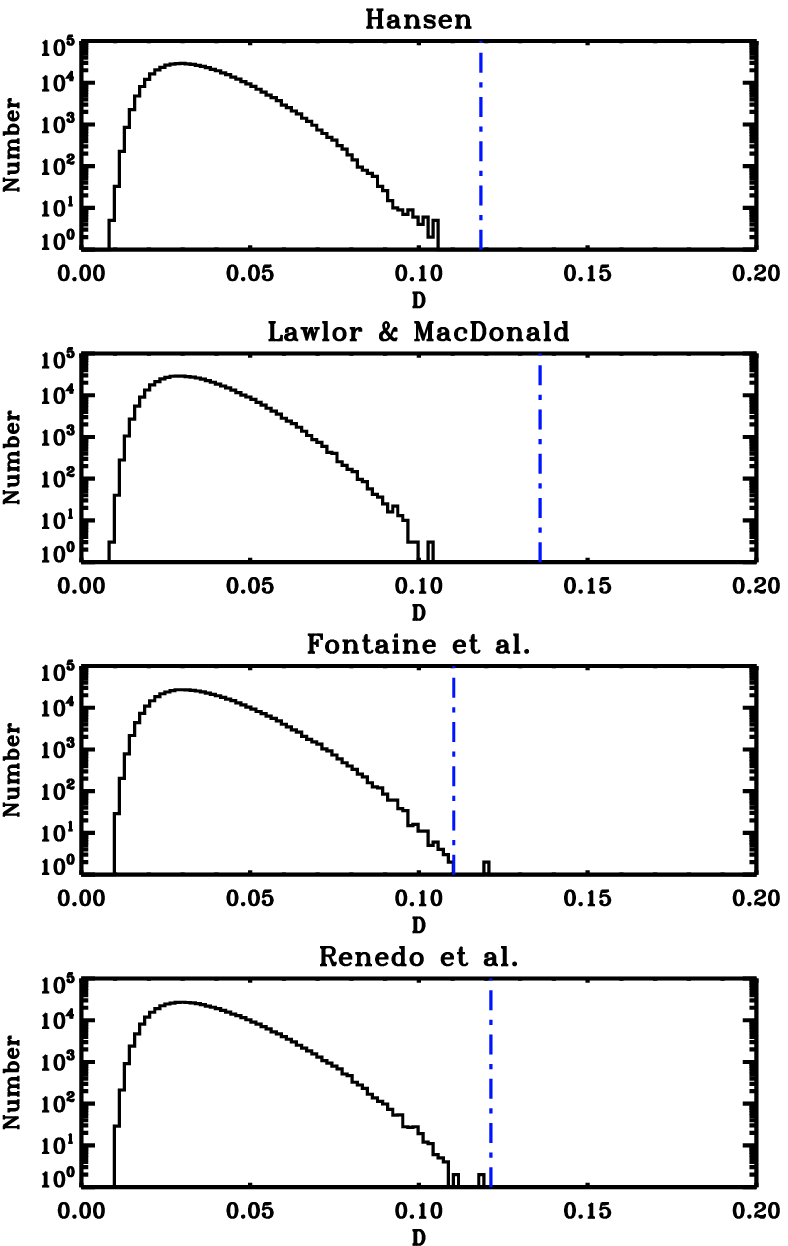}
\caption{Results from our iterative error analysis using data from WFC3.  The blue line indicates the value of D calculated when comparing our data to each model.  Our data are inconsistent with being drawn from any of these models at the 3$\sigma$ level.  This is true even after considering a conservatively large systematic uncertainty in distance and a random uncertainty in temperature.}
\label{ks_sim}
\end{figure}

\begin{figure}[h!tbp]
\centering
\includegraphics[scale=0.95]{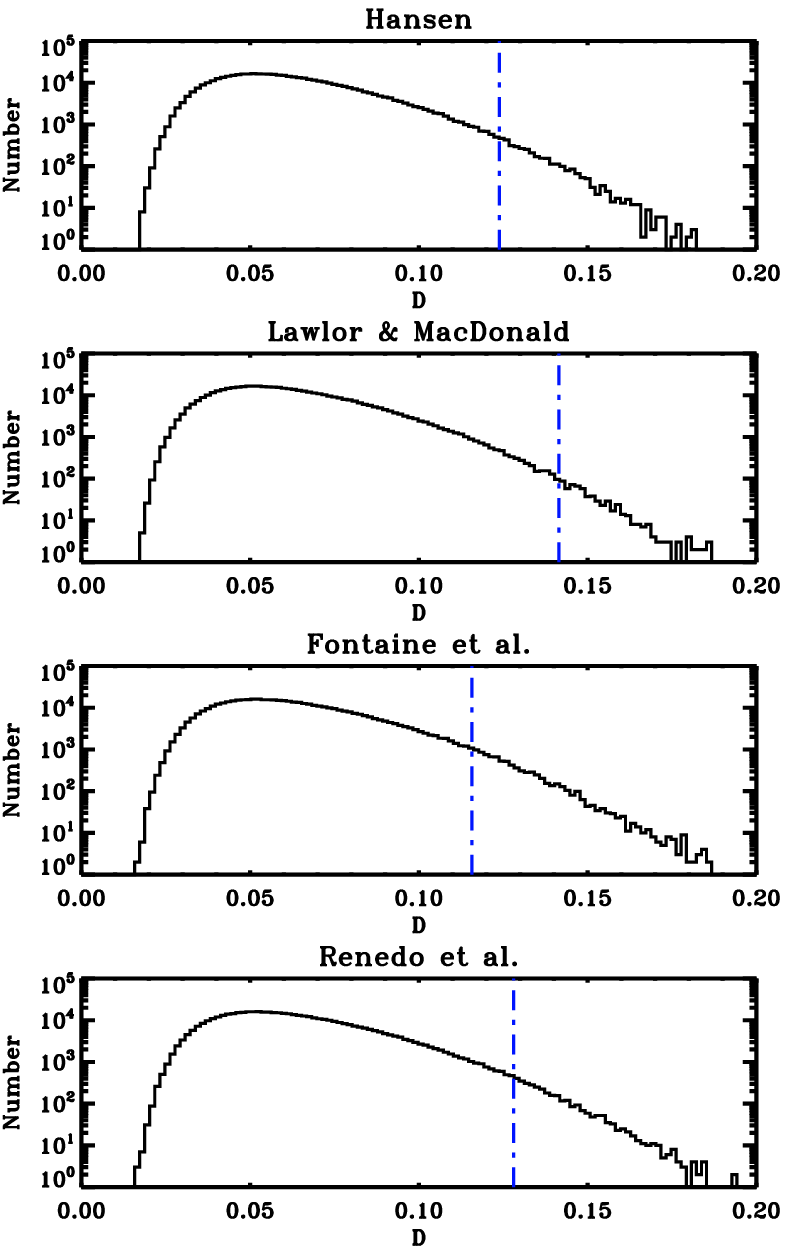}
\caption{Results from our iterative error analysis using data from the ACS field.}
\label{ks_sim_acs}
\end{figure}

From these cumulative distributions we can calculate the Kolmogorov-Smirnov (KS) separation statistic ``$D$", indicating the largest separation between the cumulative distribution of our data and that of the model.  Next, we draw samples of 813 objects from each model distribution 500,000 times.  Here we use a sample size of 813 rather than our full sample of 887 because only 813 of our objects have temperatures that fall between 7,000 K and 30,000 K.  Each time a random sample of temperatures is drawn from the model, each temperature is given a random shift corresponding to our measured temperature fit error distribution.  All of the temperatures in each sample are also given the same shift corresponding to our systematic temperature error resulting from an uncertain distance to the cluster.  The uncertainty in distance leads to a systematic uncertainty in all of our absolute magnitudes and therefore a systematic uncertainty in all of our fit temperatures.  The errors added to the model samples correspond to a 1$\sigma$ distance modulus uncertainty of 0.15.  This is a conservative estimate, and is actually 50\% greater than the standard deviation of the values quoted in \cite{woodley-2012-aj}.  In each iteration we calculate the value of D between the sample and the model it was drawn from.  We then build a distribution of these $D$ values and ask how likely it would be to draw a sample from these models that produce a value of $D$ greater than or equal to that of our data.  These distributions are shown in Figures \ref{ks_sim} and \ref{ks_sim_acs}.

Even with this added uncertainty, our data are ruled out as being consistent with any of these cooling models at around 3$\sigma$.  Hansen and Renedo et al. are consistent at roughly 3$\sigma$, Lawlor \& MacDonald at around 3.5$\sigma$, and Fontaine et al. at a bit less than 3$\sigma$.  This is due largely to the sharpness of the transition in our cooling curve, and the temperature at which the transition occurs.  None of the models show such an abrupt transition to the Mestel slope, and so the predicted cumulative distributions differ significantly from the observations.  It is important to note that the age scale determined from the giant branch in Section \ref{msrate} can only add further constraint.  This value will not affect the cumulative distributions presented above in any way, and can not make the data more consistent with the models.  

We also used other cooling models available from \citet{bergeron-2011-web}.  We carry out the same comparison described above with the mixed C/O thick atmosphere models, as well as the mixed C/O thin atmosphere models, in addition to the pure C thick atmosphere models used above.  We found that, given our sample size and temperature range, we can not distinguish between these models and our data are not consistent with any of them.  Unfortunately, from this comparison we gain no insight into the reasons for the discrepancy between the data and the models.  It appears this difference cannot be easily explained by varying the core composition or H atmosphere thickness.  

In addition, we consider the effects of varying the DB fraction on the expected cumulative distribution of our temperatures and magnitudes (our original approach assumes our sample is 100\% DA stars).  We find that we can not make the data significantly more consistent with the models of \citet{fontaine-2001-pasp} by allowing the DB fraction to vary freely.

A comparison of the temperature distribution from the ACS field to each model is shown in Figure \ref{ks_sim_acs}.  Even though the difference between the model distributions and ACS data is almost identical to the difference between the WFC3 data and the models, due to the smaller sample size (188 vs. 813) these data are consistent with all of the models at around 2$\sigma$.

We also demonstrate in Figure \ref{ks_dist} that the inconsistency between the models and our data can not be explained in terms of a significantly different distance.  We repeat our fitting procedure while assuming the two most extreme distance measurements from those summarized in Table 1 of \citet{woodley-2012-aj}.  We then build a cumulative distribution of fit temperatures and compare this to the models.  Even centering our distance prior 2$\sigma$ from our mean distance of 13.30, our results are inconsistent with any of the models at just below 3$\sigma$.

\begin{figure*}[h!tbp]
\centering
\includegraphics[scale=0.95]{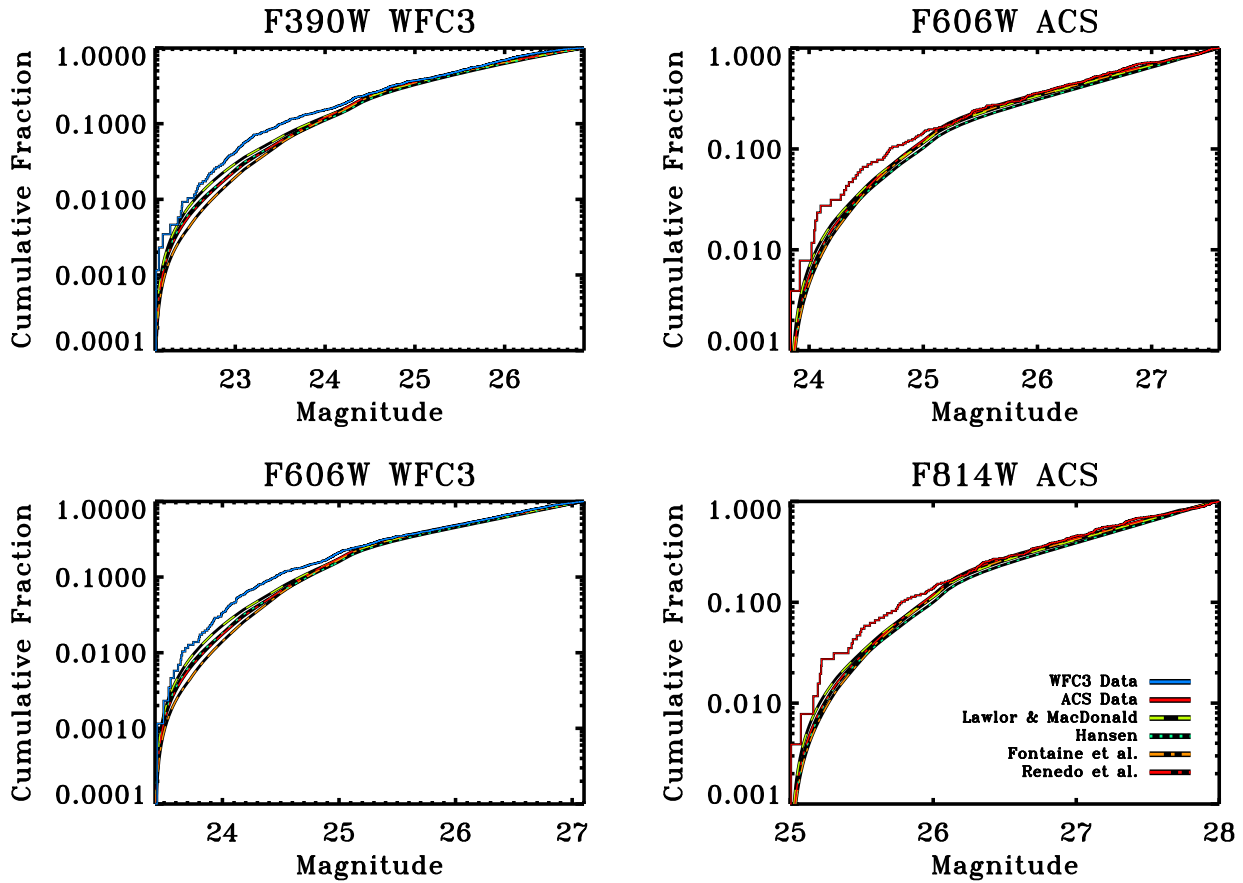}
\caption{Cumulative distributions of our object magnitudes in each filter compared with those predicted by the models.}
\label{mag_cumul}
\end{figure*}

\begin{figure}[h!tbp]
\centering
\includegraphics[scale=0.5]{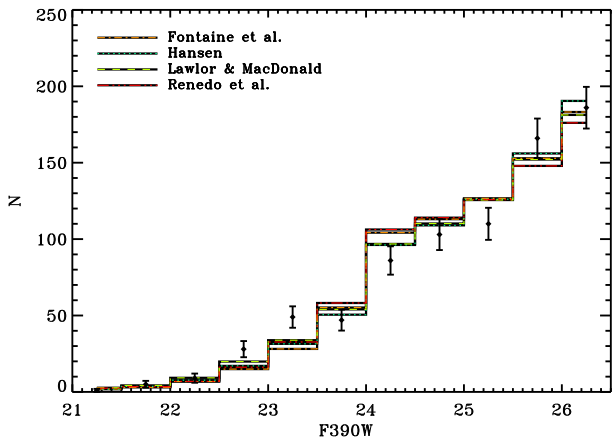}
\caption{The luminosity function of our WFC3 data compared to the models over the F390W range [ 21 , 26.5 ].}
\label{lf_compare}
\end{figure}

Finally, we perform a simple sanity check to show that the difference between the data and the models is not a result of bad data in one or more of our filters.  We calculate the expected cumulative distribution of object magnitudes in each of our four filters for each model.  We then compare this to the observed cumulative distributions.  These distributions are shown in Figure \ref{mag_cumul}. All of these distributions show the data weighted toward brighter magnitudes than the models.  This agrees with the results we find when making the comparison in temperature space.  We also show the comparison to the luminosity function in $F390W$ in Figure \ref{lf_compare}.  We also observe a discrepancy between the models and data in this space as the luminosity function is just the derivative of the cumulative distribution (or, more naturally perhaps, the cumulative distribution is the integral of the standard luminosity function as shown in Equation \ref{lf_eq}.)  In this equation the cumulative distribution is calculated over the magnitude range $[M_1,M_2]$ and normalized to 1.  The term $\Phi (M)$ is the luminosity function (dN/dM).  Calculating the cumulative distribution directly avoids the need to bin the data in either magnitude or temperature.

\begin{equation}
C(M)=\frac{\int_{M_{1}}^{M}\Phi (M') dM'}{\int_{M_{1}}^{M_{2}}\Phi (M') dM'}
\label{lf_eq}
\end{equation} 

\section{Discussion}
\label{discuss}

We have presented an empirical measurement of the cooling rate of white dwarfs in the cluster 47 Tucanae.  Our results suggest two linear regions (in log-space) with a transition occuring around 20,000K.  In the upper region, between approximately 35,000K and 20,000K, the temperature of the white dwarfs changes like $t^{-0.19}$.  This slope agrees with all of the models within uncertainties.  In the cooler region, between approximately 19,000K and 7,000K, the temperature decreases like $t^{-0.42}$.  This is close to Mestel's original value of $-0.40$ and in fact agrees precisely with more thorough modern calculations, such as those summarized in \citet{hansen-2004-pr}.  This region of the cooling curve agrees with all of the models as well.  However, there is a clear discrepancy between the abrupt transition seen in the data and the more gradual transition seen in the models.

Our measurements of both the white dwarf temperatures and their age on the sequence use a robust maximum likelihood fitting method and a thorough treatment of all the uncertainties involved.  The disagreement of our data with the available models, as well as inconsistency between the models themselves, suggests differences in the input physics involved in the transition region between the two power law regimes.  The results of our analysis can therefore be used to constrain the input physics of these models.

Such cooling models are used to predict the luminosity function of white dwarf populations of various ages, and are used to date stellar populations, such as the Galactic disk \citep{oswalt-1996-nat}, old globular clusters \citep{hansen-2007-apj}, and younger open clusters \citep{hippel-2005-apj}.  These results therefore have potentially widescale implications for the age of the Galaxy and it's various components.

Our fit temperatures from all 887 objects from the WFC3 data and 292 objects from our ACS data, as well as the age for these objects, will be made available online.

\vspace{12pt}

Support for the program GO-11677 was provided by NASA through a grant from the Space Telescope Science Institute, which is operated by the Association of Universities for Research in Astronomy, Inc., under NASA contract NAS 5-26555.  This work is supported in part by the NSERC Canada.

\pagebreak

%    5. Bibliography
\bibliographystyle{apj}
\bibliography{biblio}

\end{document}